\documentclass[aps,prl,preprint,superscriptaddress]{revtex4-1}

\usepackage{graphicx}
\begin{document}

\title{Electron-phonon coupling-induced kinks in the $\sigma$ band of graphene}
\author{Federico Mazzola}
\author{Justin W. Wells}
\email[]{quantum.wells@gmail.com}
\affiliation{Department of Physics, Norwegian University of Science and Technology (NTNU), N-7491 Trondheim, Norway }
\author{Rositza Yakimova}
\affiliation{Department of Physics, Chemistry, and Biology, Link\"oping University, S-581 83 Link\"oping, Sweden}
\author{S\o ren Ulstrup}
\author{Jill A. Miwa}
\author{Richard Balog}
\author{Marco Bianchi}
\affiliation{Department of Physics and Astronomy, Interdisciplinary Nanoscience Center (iNANO), Aarhus University, 8000 Aarhus C, Denmark.}
\author{ Mats Leandersson}
\author{ Johan Adell}
\affiliation{MAX IV Laboratory, Lund University, P.O. Box 118, 221 00 Lund, Sweden}
\author{Philip Hofmann}
\affiliation{Department of Physics and Astronomy, Interdisciplinary Nanoscience Center (iNANO), Aarhus University, 8000 Aarhus C, Denmark.}
\author{ T. Balasubramanian}
\affiliation{MAX IV Laboratory, Lund University, P.O. Box 118, 221 00 Lund, Sweden}

\date{\today}

\date{\today}
 \begin{abstract}
Angle-resolved photoemission spectroscopy reveals pronounced kinks in the dispersion of the $\sigma$ band of graphene. Such kinks are usually caused by the combination of a strong electron-boson interaction and the cut-off in the Fermi-Dirac distribution. They are therefore not expected for the $\sigma$ band of graphene that has a binding energy of more than $\approx 3.5$~eV. We argue that the observed kinks are indeed caused by the electron-phonon interaction, but the role of the Fermi-Dirac distribution cutoff is assumed by a cut-off in the density of $\sigma$ states. The existence of the effect suggests a very weak coupling of holes in the $\sigma$ band not only to the $\pi$ electrons of graphene but also to the substrate electronic states. This is confirmed by the presence of such kinks for graphene on several different substrates that all show a strong coupling constant of $\lambda \approx 1$. 

PACS numbers: 81.05.ue,  73.22.Pr,   63.70.+h

 \end{abstract}
 \maketitle
Many-body interactions can strongly affect the spectral function of solids and their presence is frequently heralded by  so-called kinks in the dispersion of the electronic states near the Fermi energy, as observed by angle-resolved photoelectron spectroscopy (ARPES). Such kinks are primarily caused by electron-boson interactions and many cases of electron-phonon \cite{Hengsberger:1999b,Lashell:2000,Gayone:2005} or electron-magnon \cite{Schafer:2004} induced kinks have been reported. In the cuprate high-temperature superconductors, strong kinks have been found near the Fermi energy \cite{Johnson:2001,Lanzara:2001} and their origin as well as their significance for the mechanism for high-temperature superconductivity has given reason to some debate \cite{Anderson:2007}. The observed kinks in the spectral function contain a wealth of information about the underlying many-body interactions, such as the strength of the coupling as a function of position on the Fermi surface, as well as the energy of the bosons \cite{Bianchi:2010}. Note, however, that the presence of kinks does not necessarily imply the presence of bosonic interactions in correlated materials \cite{Byczuk:2007}. 
 
The observation of a kink signals a strong change in the real part of the self-energy $\Sigma'$ that describes the deviation of the observed dispersion from the single-particle picture \cite{Hofmann:2009b}. The origin of this structure can most easily be understood by considering the imaginary part of the self-energy $\Sigma''$ that is inversely proportional to the lifetime of the ARPES photohole and related to $\Sigma'$ via a Kramers-Kronig transformation. Far away from the Fermi energy $E_F$, a photohole can be filled by electrons from lower binding energies dropping into the hole, emitting a boson of energy $\hbar \omega_E$ to conserve energy and momentum. For binding energies smaller than $\hbar \omega_E$  this is no longer possible, leading to a marked increase in lifetime. The corresponding decrease in $\Sigma''$ leads to a maximum in $\Sigma'$ and this gives rise to the kink \cite{Hofmann:2009b}.

\begin{figure}
\centering
\includegraphics [width=0.6\columnwidth] {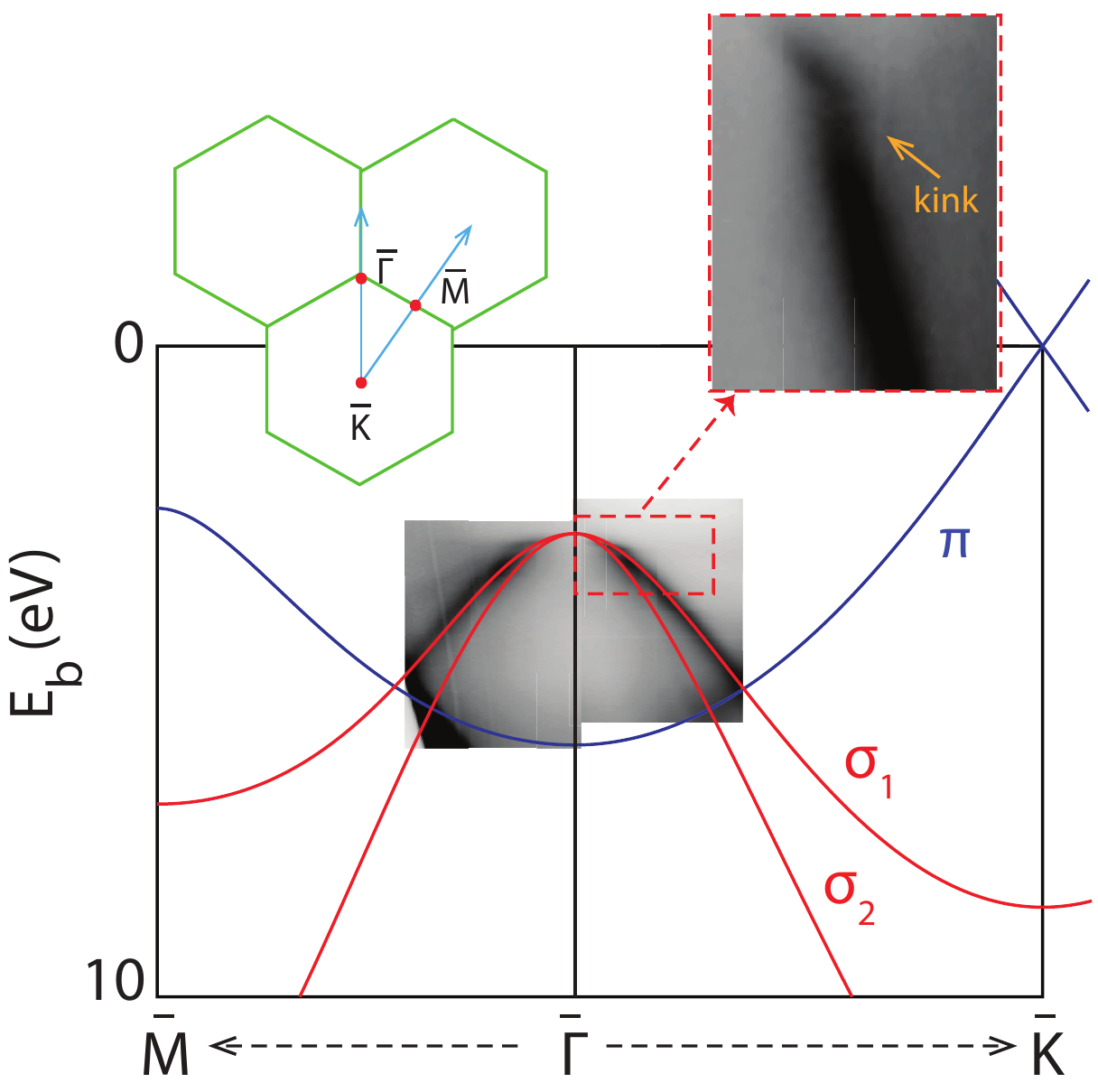}
\caption{(color online) Non-interacting (tight-binding) band structure of graphene, depicting the $\sigma$ bands (red) and the $\pi$-band (blue) \cite{Reich:2002}. The Brillouin zone is depicted in the inset. The $\sigma$ band  consists of two branches, $\sigma_{1}$ and $\sigma_{2}$, meeting at a common maximum at $\overline{\Gamma}$, with a binding energy of 3.5 to 4.0~eV. ARPES data for MLG graphene on SiC (greyscale) are superimposed. The detailed dispersion in the vicinity of the $\sigma$ band maximum in the $\overline{\Gamma}-\overline{\textrm{K}}$ direction is magnified. The measured dispersion deviates from the non-interacting behaviour, showing a clear kink accompanied by a band narrowing near the top of the $\sigma$ band. The photon energy and temperature of data acquisition were $h\nu=36$~eV and $T=100$~K respectively.}
\label{overview}
\end{figure}

\begin{figure}[b]
\includegraphics [width=0.6\columnwidth] {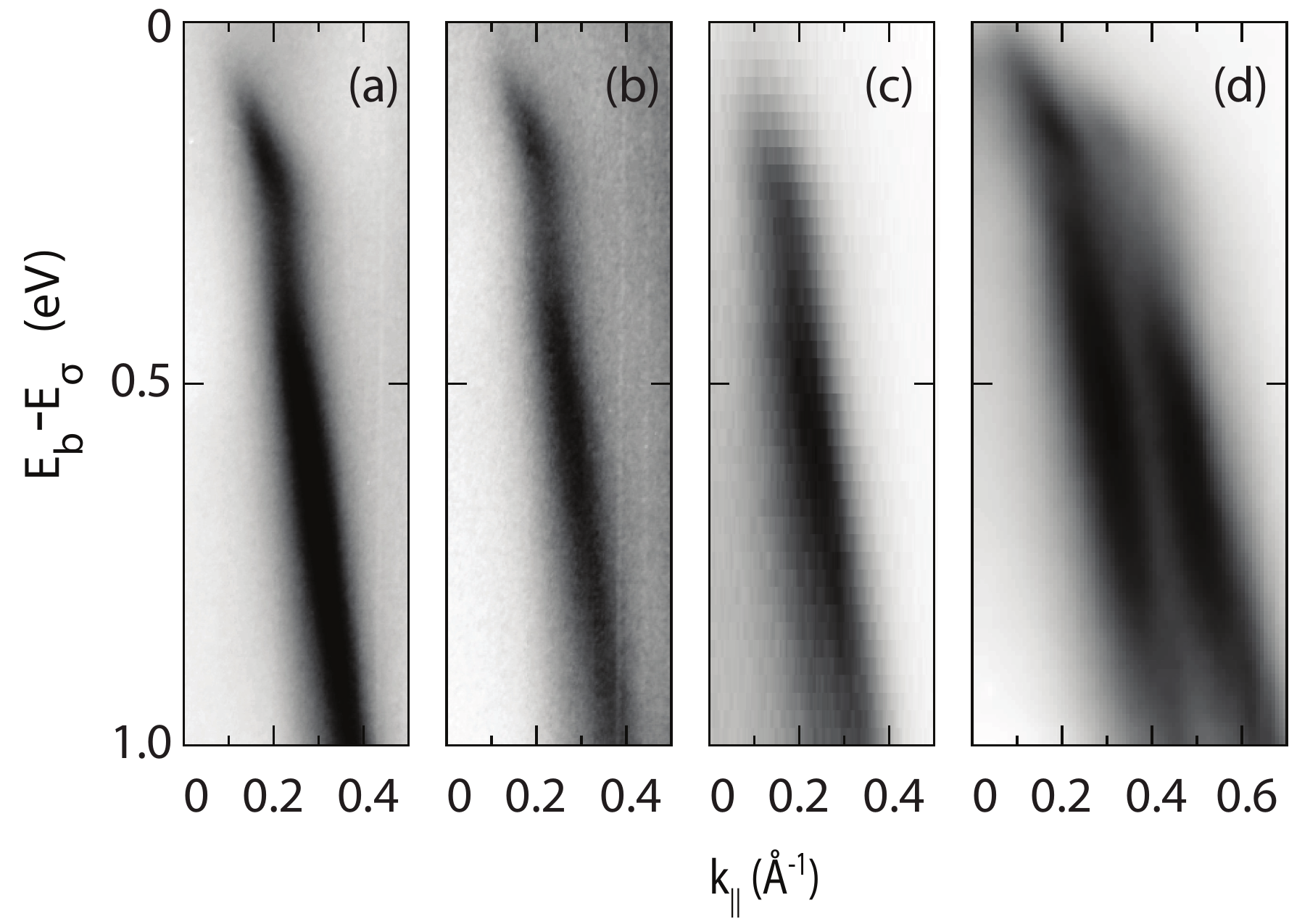}
\caption{ARPES data acquired for (a) quasi-freestanding-monolayer graphene on Ir(111), (b) Rb-doped quasi-freestanding-monolayer graphene on Ir(111), (c) monolayer graphene on SiC and (d)  bilayer graphene on SiC. The energy scales are plotted relative to the $\sigma$ band maximum ($E_{\sigma}\approx 3.5$ ~eV). The dispersion direction  corresponds to the  $\overline{\Gamma}-\overline{\textrm{K}}$ for all the three systems.  The photon energy of data acquisition was $h\nu=36$~eV} 
\label{comparison}
\end{figure}

The lack of occupied states above $E_F$ (at low temperature) is thus crucial for the appearance of the kink and many-body effect related dispersion kinks are thus only expected near $E_F$, at least for the coupling to bosonic modes. In this Letter, we report the observation of pronounced kinks near the top of the $\sigma$ band in graphene/graphite (see Figs.\  \ref{overview} and \ref{comparison}). Since these states are found at a binding energy of $>3.5$~eV, the presence of such kinks is unexpected. We show that the observed spectral features can still be explained by a strong electron-phonon interaction but the role of the Fermi-Dirac distribution cutoff is assumed by the density of  $\sigma$ states. This novel mechanism suggests that the hole in the $\sigma$ band primarily decays through electrons from the same band instead of electrons from the $\pi$ band or the substrate. This is confirmed by the observation of a similarly strong coupling for a large variety of graphene systems. 

ARPES data were collected for six different material systems at three different synchrotron radiation beamlines: graphite, epitaxial monolayer (MLG) and bilayer (BLG) graphene on SiC \cite{Virojanadara:2008} at beamlines I3 \cite{Balasubramanian:2010} and I4 \cite{Jensen:1997} of MAX-III, as well as  MLG graphene on Ir(111), oxygen-intercalated quasi-free standing monolayer graphene (QFMLG) on Ir(111), with and without Rb doping, on the SGM-3 line of ASTRID \cite{Hoffmann:2004}. Measurements were carried out under ultrahigh vacuum and temperatures which are low  (see Table \ref{tab1}) compared to those required for the excitation of optical phonon modes. The energy and momentum resolutions were better than $35$~meV and $0.01$~\AA$^{-1}$, respectively. 

Figs.\ \ref{overview} and \ref{comparison} illustrate the strong renormalization of the $\sigma$-band for different graphene systems. An overview is given in Fig.\ \ref{overview} showing the non-interacting (tight-binding) band structure of graphene \cite{Reich:2002} together with the ARPES data for  MLG graphene on SiC (acquired at 100~K) near the top of the $\sigma$ band. The general agreement of data and calculated band structure is satisfactory. However, a closer inspection shows the formation of a pronounced kink in the dispersion near the top of the band, accompanied by a band narrowing, the characteristic sign of a strong electron-boson interaction. While such kinks are expected and observed for doped graphene near the Fermi energy \cite{Bostwick:2007, Bostwick:2007b, Bianchi:2010}, their appearance at a high binding energy of $\approx$~3.5~eV is unexpected. Results for graphene and bilayer graphene show that the strong renormalization is an ubiquitous feature (see Fig.\ \ref{comparison}). The energy scale in this figure has been defined relative to the top of the $\sigma$ band extrapolated from the high energy dispersion of  the bands. For these graphene systems, matrix element effects can strongly suppress the photoemission intensity of one or both of the $\sigma$ bands at the chosen photon energy and experimental geometry \cite{Shirley:1995b}. Indeed, at normal emission, either  it is only possible to see a single branch of the two forming the $\sigma$ band or the intensity of one branch is drastically reduced compared to the other  (see Fig.\ \ref{comparison}).

While the electron-phonon coupling appears to be an obvious candidate for the appearance of the kinks, the mechanism must be very different from the situation near the Fermi level where the Fermi-Dirac function cutoff is ultimately responsible for the strong change in the self-energy. Consider the imaginary part of the self energy $\Sigma''$ for the electron-phonon coupling \cite{Grimvall:1981,Hofmann:2009b}:
\begin{eqnarray}
\Sigma''\big(
\epsilon _i,T \big)= \pi\int\limits_{0}^{\omega_{\mathrm{max}}}\{\alpha
^{\mathrm{2}} F^{\mathrm{A}}(\epsilon _i,\omega) [1 +
n(\omega) - f(\epsilon _i-\omega)] \nonumber\\
+ \alpha^{\mathrm{2}} F^{\mathrm{E}}(\epsilon_i,\omega)
[n(\omega) + f(\epsilon _i+\omega)]\} \mathrm{d}\omega\  +\Sigma''_0.
\label{life-full}
\end{eqnarray}
where $\alpha^2F^{A(E)}$ are the Eliashberg coupling functions for phonon absorption (emission), $\epsilon_i$ is the initial state energy of the hole with respect to the top of the $\sigma$ band and $\omega$ is the phonon energy. $n$ and $f$ are Bose-Einstein and Fermi-Dirac distributions respectively and $\Sigma''_0$ accounts for electron-defect and electron-electron scattering, which is assumed to be independent of $\epsilon_i$ in the small energy range of interest here. Far from $E_F$, as in the present situation, we can approximate $f=1$. If we assume that the electron-phonon interaction is dominated by an optical Einstein mode of $\hbar \omega_E \approx 190$~meV, we further find that $n(\omega_E) \ll 1$. The  expression simplifies to 
\begin{eqnarray}
\Sigma''\big(
\epsilon _i,T \big)=\pi\int\limits_{0}^{\omega_{\mathrm{max}}} \alpha^{\mathrm{2}} F^{\mathrm{E}}(\epsilon_i,\omega) \mathrm{d}\omega\  +\Sigma''_0.
\label{life-simple}
\end{eqnarray}
The crucial point now is that the Eliashberg function is strongly energy dependent: if we only consider electron-phonon scattering events within the $\sigma$ band, a hole that is closer than $\hbar \omega_E$ to the top of the $\sigma$ band cannot decay by the emission of an optical phonon but a hole at a slightly larger binding energy can. More precisely, the phase space for the electron-phonon scattering is given by the density of states in the $\sigma$ band. For a two-dimensional (nearly) parabolic band, this density of states is well-approximated by a step function. Hence, we can write the Eliashberg function as
\begin{eqnarray}
\alpha^{\mathrm{2}} F^{\mathrm{E}}(\epsilon_i,\omega) = \frac{\omega_E}{2} \lambda \delta(\omega - \omega_E) \Theta(\epsilon_i - \omega_E),
\label{eliash}
\end{eqnarray}
where $\lambda$ is the electron-phonon coupling constant and $\Theta$ the Heaviside function. This corresponds to the standard model Eliashberg function for coupling to an Einstein mode but the mechanism is only permitted for $\epsilon_i > \omega_E$. 

\begin{figure}
\centering
\includegraphics [width=0.6\columnwidth] {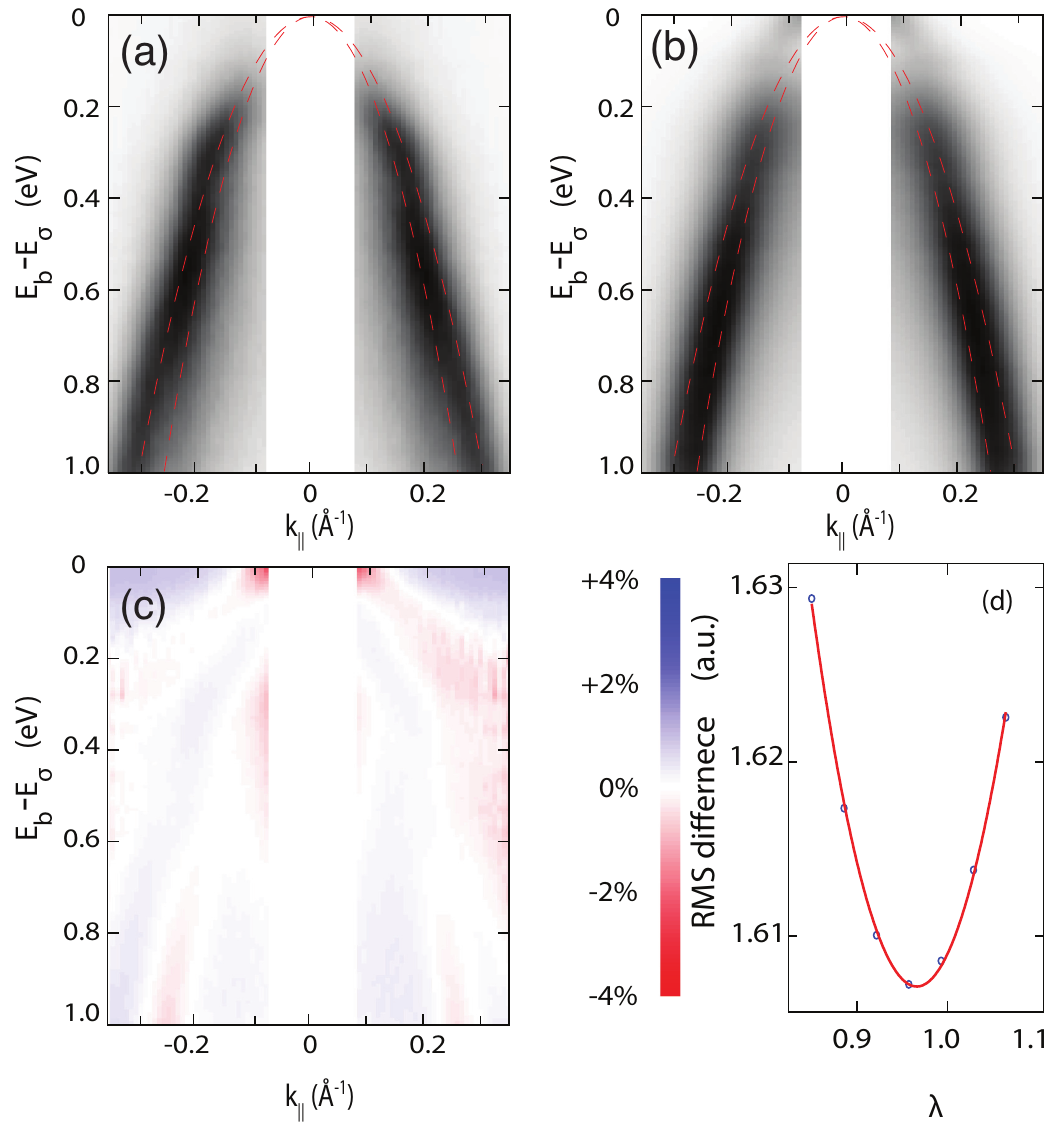}
\caption{(a) ARPES data for the $\sigma$ band of MLG graphene on SiC along the $\overline{\textrm{K}}-\overline{\Gamma}-\overline{\textrm{K}}$ direction and (b) model spectral function derived from equations (\ref{life-simple}) and (\ref{eliash}). The binding energy is shown relative to the $\sigma$ band maximum, ($E_\sigma\approx 3.5$~eV).  The red parabolae depict the expected dispersion in a non-interacting model. (c) Difference between (a) and (b). (d) Sum of the root mean square difference of the pixels in (a) and (b) as a function of $\lambda$, optimising all the other parameters in the model for each $\lambda$ value.  The photon energy of data acquisition was $h\nu=36$~eV and the temperature 100~K. }\label{meas_sim}
\end{figure}

We test this model by using it to calculate the spectral function and compare it to the experimental data. This is illustrated in Fig.\ \ref{meas_sim} for MLG  on SiC at low temperature ($T=100$~K). If we assume that $\hbar \omega_E=190$~meV, the only free parameters in the model are $\lambda$, $\Sigma''_0$ and those describing the parabolic bare band dispersion. The Einstein energy mode ($\hbar \omega_E=190$~meV) is fixed such that the final reconstructed band reaches the best agreement with the real spectra giving a further confirmation of the energy scale for kinks. From $\Sigma''$ we obtain $\Sigma'$ by a Kramers-Kronig transformation. The bare dispersion of the $\sigma$ band is approximated by two parabolae (one parabola describing each branch). 

 For the comparison with the calculated spectral function, the measured data undergo a background subtraction and an intensity normalisation (such that the measured geometry-induced difference in matrix elements between the $+k_\parallel$ and $-k_\parallel$ directions is averaged).  The region $[-0.08\le k_\parallel \le0.08]$~\AA$^{-1}$, is excluded since the matrix elements are so small that the ARPES intensity approaches zero \cite{Shirley:1995b}. The simulated spectral function is convolved by experimental energy and momentum resolutions, and normalised to the same intensity as the measurement.  The agreement between measured and calculated spectral function shown in Fig.\  \ref{meas_sim}(a) and (b) is quantified by the sum of the root mean square (RMS) differences between the pixel values in data and model. The parameters in the simulation ($\lambda$, $\Sigma''_0$ and those describing the bare dispersion) are optimised until the lowest sum of RMS differences is reached. Fig.\  \ref{meas_sim}(c) shows that the difference between model and data is very small, at most a few \%. Fig.\  \ref{meas_sim}(d) shows the sum of RMS differences as a function of $\lambda$ with all the other parameters optimized for each $\lambda$ value. For the present case of MLG graphene on SiC, we find $\lambda=0.96\pm0.04$.

The spectral function derived from our simple model thus provides and excellent description of the data but it is worthwhile to discuss the assumptions behind it: the observation of kinks is only possible if the density of states cutoff replaces the usual cutoff of the Fermi-Dirac function. This, in turn, relies crucially on the fact that holes in the $\sigma$ band are not filled by electrons from the (degenerate) $\pi$ band or by electrons from the substrate (SiC, Ir) or nearby atoms (O, Rb). This appears to be a reasonable assumption since the substrate bonding is primarily mediated through the $\pi$ electrons with little involvement of the $\sigma$ band. It can also be tested experimentally: if valid, one would expect the $\sigma$ band kink to be ubiquitous and of similar strength in all graphene-based systems. We have therefore collected and analyzed data from the different graphene systems. The results are given in Table \ref{tab1}. As already seen in Fig.\ \ref{comparison}, the effect appears to be present independently of the substrate, decoupling by intercalation or electron doping. The coupling strength $\lambda$ is also remarkably similar across the samples studied, although slightly smaller for BLG/SiC, indicating that interlayer interactions probably play a small role.

\begin{table}
\caption{Results of the analysis carried out for different graphene systems. The electron-phonon coupling parameter $\lambda$ and its uncertainty are quoted, together with the sample temperature ($T$) during acquisition of ARPES data.}
\begin{tabular}{| c | c | c |} \hline\hline
\textbf{System} & \textbf{$\lambda \pm \Delta \lambda$} & \textbf{$T$ (K)}\\ \hline
MLG/SiC & $ 0.96 \pm 0.04$  & 100\\  \hline
MLG/SiC & $ 0.97 \pm 0.04$ & 300\\ \hline
BLG/SiC &$0.75\pm0.05$  & 100 \\ \hline
Graphite & $ 0.97 \pm 0.04$  & 100\\  \hline
MLG/ Ir & $ 0.97 \pm 0.05$  & 70 \\  \hline
QFMLG/Ir  & $ 0.96 \pm 0.04$  & 70 \\  \hline
QFMLG/Rb/Ir & $ 0.96 \pm 0.04$  & 70 \\ \hline \hline
\end{tabular}\label{tab1}
\end{table}

The values of $\lambda$ reported in Table \ref{tab1} are not only similar, they are also high on an absolute scale, in the same order of magnitude as for a strong coupling BCS superconductor \cite{Grimvall:1981}. This is in contrast to the $\pi$ band where very small $\lambda$ values have been found near $E_F$ for the weakly p-doped case \cite{Johannsen:2013,Ulstrup:2012}, and somewhat stronger coupling upon electron doping  \cite{Bostwick:2007,Bostwick:2007b,Bianchi:2010}. Note, however, that even though the density of states is zero both at the Dirac point and at the top of the $\sigma$ band, its energy dependence is very different. It linearly increases for the $\pi$ band but is a step function for the $\sigma$ band. This step function is ultimately responsible for the observation of the kink because it instantaneously changes the coupling strength from zero to the high $\lambda$ observed here. Energy-dependent changes of $\lambda$ have been observed before \cite{Gayone:2003} but in most (three dimensional) systems the changes in the density of states causing them are more gradual than here.

The strong coupling and the high phonon energies could potentially give rise to a high transition temperature for superconductivity $T_c$. However, being so far from the Fermi level, the $\sigma$ states do not play any role in conduction. This is in sharp contrast to the closely related system, MgB$_2$, whose quasi-two dimensional $\sigma$ band does cross the Fermi level and  is mainly responsible for its large $\lambda$ and $T_c$ of about 40~K \cite{Mazin:2003}. One can only speculate that by substitutional doping of graphene, one could push the sigma band to the Fermi level and preserve the large $\lambda$  and obtain a high $T_c$ \cite{Singh:2002,Blase:2009}.

In conclusion, we have observed  an electron-phonon coupling-induced kink near the top of the $\sigma$ band of graphene. The kink is placed far away from the Fermi level and cannot be explained by the cutoff in the Fermi-Dirac function. Instead, its observation is ascribed to the quasi-instantaneous change in the density of states of the $\sigma$ band. The electron-phonon coupling is found to be strong ($\lambda \approx 1$) and the kink is ubiquitous for graphene systems. Its presence suggests that the $\sigma$ band is decoupled not only from the $\pi$ states but also from the electronic states of the substrate. This is not unexpected but it also suggests that the strength of the observed kink can provide information of the interaction between graphene and its surroundings.

We gratefully acknowledge financial support from the Lundbeck foundation, the VILLUM foundation, The Danish Council for Independent Research / Technology and Production Sciences. The work carried out at the MAX IV laboratory was made possible through support from the Swedish research council and Knut and Alice Wallenberg Foundations. T.B acknowledges L. Walld{\'e}n for his encouragement and discussions.

\vspace{-0.2in}
%




\end{document}